\documentclass{emulateapj}
\usepackage{natbib}
\usepackage{amsfonts,amsmath,amssymb}

\shorttitle{Recollimation boundary layers in CTTS}
\shortauthors{H. M. G{\"{u}}nther et al.}

\begin{document}


\title{Recollimation boundary layers as X-ray sources in young stellar jets}


\author{Hans Moritz G{\"{u}}nther}
\affil{Harvard-Smithsonian Center for Astrophysics, 60 Garden Street,
  Cambridge, MA 02139, USA}
\email{hguenther@cfa.harvard.edu} 
%\and
\author{Zhi-Yun Li}
\affil{Department of Astronomy, University of Virginia, P.O. Box 400325, Charlottesville, VA 22904, USA} 
%\and
\author{P. C. Schneider}
\affil{Hamburger Sternwarte, Gojenbergsweg 112, 21029, Hamburg, Germany}

\begin{abstract}
Young stars accrete mass from circumstellar disks and in many cases, the
accretion coincides with a phase of massive outflows, which can be
highly collimated. Those jets emit predominantly in the optical and IR
wavelength range. However, in several cases X-ray and UV observations
reveal a weak but highly energetic component in those jets. X-rays are
observed both from stationary regions close to the star and from knots
in the jet several hundred AU from the star. In this article we show
semi-analytically that a fast stellar wind which is recollimated by the
pressure from a slower, more massive disk wind can have the right
properties to power stationary X-ray emission. The size of the shocked
regions is compatible with observational constraints. Our calculations
support a wind-wind interaction scenario for the high energy emission
near the base of YSO jets. For the specific case of DG~Tau, a stellar
wind with a mass loss rate of
$5\cdot10^{-10}\;M_{\odot}\;\mathrm{ yr}^{-1}$ and a wind speed of
800~km~s$^{-1}$ reproduces the observed X-ray spectrum. We conclude that
a stellar wind recollimation shock is a viable scenario to power
stationary X-ray emission close to the jet launching point.

\end{abstract}

\bibliographystyle{apj}



\bibliographystyle{aa}

\section{Introduction} 
In many areas of astrophysics compact central objects accrete mass and angular momentum from a disk and at the same time they eject a highly collimated jet. This is seen for central objects as massive as AGN or as light as (proto) brown dwarfs. Jets from AGN or accreting neutron stars reach relativistic velocities while the material in jets from young stellar objects (YSOs) is significantly slower. 
A large number of jets is observed in nearby star forming regions, where the jet composition and structure can be studied in great detail \citep[see the review by][]{2014arXiv1402.3553F}.
Jet launching starts in the early stages of star formation and continues until the accretion from the circumstellar disk cedes. Jets from very young stars are the most powerful and can sometimes be traced up to several parsec from the source. In these systems, the central engine is still deeply embedded in a dense envelope of gas and dust and thus cannot be observed directly. As YSOs evolve, the envelope becomes thinner. Actively accreting low-mass stars in this stage are called classical T Tauri stars \citep[CTTS -- for a review see][]{2013AN....334...67G}. Their jets often only reach a few hundred AU (and are thus sometimes called ``microjets'' in comparison to the outflows from younger objects).
It seems reasonable to suspect that the same physics governs the launching from any type of young star and that the same processes occur close to the launch site, but observationally the inner few tens of AU are only accessible in CTTS, where we can study the initial properties before the jet interacts with the ambient medium. 

Recently, there has been increasing evidence that jets from CTTS have a stationary, hot (in the MK range), X-ray emitting region only a few tens of AU from the central star (Section~\ref{sect:introxray}). In this article we want to establish recollimation boundary layers between the stellar wind and a disk wind as a viable scenario to explain stationary X-ray and UV emission from YSO jets.
While X-ray emission has been discussed in the literature (Section~\ref{sect:intromodel}), stationary shocks between a stellar wind and a recollimating disk wind have not been investigated in detail. 

X-rays trace the fastest and most energetic components of YSO jets. They can also influence the chemistry deep in the disk \citep[e.g.][]{2010ApJ...714.1511H,2012ApJ...756..157G} and thus alter the environment of planet formation because they penetrate deeper than UV and optical radiation. Unlike stellar X-ray emission, the radiation from the jet originates above the plane of the disk and thus reaches the entire disk surface, while stellar radiation may be shadowed by the inner disk rim.

In the remainder of the introduction we review observational properties (including X-ray emission) of jets from CTTS and summarize theoretical explanations for this emission in the literature. In section~\ref{sect:model} we develop the equations that govern a standing shock front and discuss the physical parameters in section~\ref{sect:parameters}. In section~\ref{sect:results} we present our results and discuss implications in section~\ref{sect:discussion}. We summarize this work in section~\ref{sect:summary}.

\subsection{Observational properties of jets from young stars}
\label{sect:introjetobs}
Outflows from CTTS consist of several layers with different flow velocities, densities, temperatures, chemical compositions and different launching positions. Conceptually, these can be thought of as conical shells, stacked into each other like the layers of an onion \citep{2000ApJ...537L..49B}. Observationally, it is not clear if there is a smooth transition between those layers or if they are separated by discontinuities. 

The slowest velocities are observed in molecular lines with typical line shifts of only a few km~s$^{-1}$ \citep{2008ApJ...676..472B}. These molecular outflows have wide opening angles around 90$^{\circ}$ \citep[e.g.][]{2013A&A...557A.110S,2014A&A...564A..11A} and are presumably launched from the disk. Faster components are seen in H$\alpha$ or in optical forbidden emission lines such as [\ion{O}{1}] or [\ion{S}{2}]. \citet{2000ApJ...537L..49B} observed the jet from the CTTS \object{DG Tau} with seven long-slit exposures of \emph{HST}/STIS to resolve the kinematic structure of the jet both along and perpendicular to the jet axis. They find that the inner, most collimated jet component moves fastest and surrounding layers have progressively lower velocities away from the jet axis. The fastest velocities seen in optical emission lines are typically 200-300~km~s$^{-1}$ \citep{2004Ap&SS.292..651B,2008ApJ...689.1112C,2013A&A...550L...1S}.

At some distance from the star, shock fronts, the so called Herbig-Haro (HH) objects, are observed when the jet runs into the ambient medium or when material emitted at higher velocities catches up with previously emitted slower material. \citet{2006A&A...456..189P} studied emission line ratios in jets in the Orion and Vela star forming regions. In their sample the knots are typically a few hundred AU from the central star. They find electron densities around $n_e \approx 50-300 \textrm{\;cm}^{-3}$ when looking at optical emission lines, slightly higher values for forbidden emission lines like [\ion{Fe}{2}] and $n_e \approx 5\times10^5-5\times10^6 \textrm{\;cm}^{-3}$ from Ca lines. The ionization fraction in HH objects is low, so the actual particle number density is one to two orders of magnitude higher. \citet{2004ApJ...609..261H} observed the CTTS \object{HN Tau} and resolved the jet at only 30~AU, much closer to the star. They find densities around $n_e=10^6-10^7\textrm{\;cm}^{-3}$ from [\ion{Fe}{2}] lines. This shows that the inner layers of jets can have densities significantly above those of the interstellar medium.

\subsection{X-ray emission from stellar jets}
\label{sect:introxray}
In some jets there is evidence for another, hotter and faster component. X-rays were first seen in the jet \object{HH 2} \citep{2001Natur.413..708P,2012A&A...542A.123S}, where the central star is invisible. Later, X-ray emission was also discovered from less enbedded CTTS, \object{RY Tau} \citep{2014ApJ...788..101S} and, most notably, \object{DG Tau}. DG~Tau is the best case to study X-ray emission from the jet close to the star for two reasons: (i)~No other jet driving young stellar object has been observed as often in X-rays. DG~Tau was the target of several shorter \emph{Chandra} exposures in 2004, 2005, and 2006 and a large program in 2010 \citep{2005ApJ...626L..53G,2008A&A...478..797G,2011ASPC..448..617G} and has been observed with \emph{XMM-Newton} in 2004 \citep{2007A&A...468..353G} and in 2012 \citep{SchneiderDGTauXray}. (ii)~DG~Tau itself is hidden behind a column density of $N_{\textrm{H}}=2\times10^{22}\textrm{ cm}^{-3}$ \citep{2008A&A...478..797G}, which absorbs any soft X-ray emission from the central star. Hard, coronal emission pierces through the gas and allows us to pinpoint the stellar position to high accuracy, while the soft X-rays observed close to the stellar position must come from the jet.

We can distinguish three different X-ray emitting regions in the DG~Tau system: First, hard emission from the central star is observed with stellar flares as seen on many other young and active stars. Second, weak and soft emission from the jet is resolved several hundred AU from the star itself. Third, additional soft X-rays are emitted close to, but not from the star, because they are subject to a much smaller absorbing column density than the central, coronal source. The centroid of the spatial distribution of soft X-rays is consistent with a position on the jet axis 30-40~AU from the star \citep{2008A&A...488L..13S,2011ASPC..448..617G} in every epoch. Thus, the jet X-ray emission appears stationary in contrast to the moving, lower temperature jet material.  The temperature of this inner emission region is remarkably stable over one decade between 3 and 4~MK; the maximum change observed is about 25\,\%. The change in luminosity is 1.6 in the same time range ($L_X=1-2\times10^{30}\textrm{ erg s}^{-1}$) \citep{SchneiderDGTauXray}.

There is no reason to believe that the DG~Tau system represents exceptional physical conditions for jet launching. While the inclination and absorption are less favorable to observe X-ray emission very close to the star for most other CTTS systems, there are indications that \object{HH 154} also shows an inner, stationary X-ray component and additional emission in the knots \citep{2011A&A...530A.123S,2011ApJ...737...54B} and that the X-ray emission in the more massive Herbig Ae/Be star \object{HD 163296} is extended in the direction of the jet by a few dozen AU, too \citep{2005ApJ...628..811S,2009A&A...494.1041G,2013A&A...552A.142G}.

In \citet{2009A&A...493..579G} we showed that the soft X-ray emission close to DG~Tau can be explained by shock heating of a jet component moving with 400-500~km~s$^{-1}$. The mass flux in this component is less than $10^{-3}$ of the total mass flux in the jet or even lower if the same material is reheated in several consecutive shocks. If the density in the fast outflow is $>10^5$~cm$^{-3}$ then the cooling length of this shock is only a few AU and it would be invisible in current optical and IR observations, since it would be surrounded and outshone by the more luminous emission from more massive, but slower jet components. However, the stationary nature of the X-ray emission was not addressed in that article. Other models for the high-energy emission for stellar jets are discussed in the next section.

\subsection{Models for high-energy emission from stellar jets}
\label{sect:intromodel}
When \citet{2001Natur.413..708P} discovered X-ray emission in HH~2, they immediately discussed strong shocks in the outflow as the most likely heating process and the most obvious way to generate a strong shock is a bow shock at the head of the outflow, where the jet encounters circumstellar material. Depending on the density contrast between the jet and the ambient material, either the forward shock that is driven into the ISM or the reverse shock that travels against the flow in the jet possesses higher shock velocities. However, in most YSO jets, we see the X-ray emission fairly close to the source, where the ISM has been cleared by outflow activity a long time ago. \citet{2003ApJ...584..843B} study the X-ray emission from \object{L1551 IRS 5}, a binary protostar. They suggest several classes of models. First, the observed X-rays might be stellar emission, that is scattered into our line-of-sight by circumstellar material. This requires high densities to reach the required scattering efficiency and the material must be neutral, otherwise it would show up as bright radio emission, which is not seen coincident with the X-ray detection in L1551~IRS~5 \citep{2003ApJ...584..843B}. In DG~Tau this scenario can be excluded because the lightcurve for the soft X-ray emission is flat, while coronal flares are seen in the hard-X-rays \citep{2011ASPC..448..617G}. If the soft X-rays were scattered stellar emission, they should display the same lightcurve.

Second, a time variable launching velocity would cause shocks to propagate along the jet, continuously heating the jet material they encounter \citep[e.g.][]{2010A&A...511A..42B,2010A&A...517A..68B}. This scenario is confirmed for an optical knot in the jet of HD~163296 \citep{2013A&A...552A.142G}, but the velocity differences are too small to heat the jet to X-ray emitting temperatures. Third, fast shocks form when the jet encounters an obstacle. In the absence of dense ISM this could be a collimation shock, the disk, or the outflow of another star. Jets are typically ejected roughly perpendicular to the disk \citep[e.g.][for IRS 5]{2002A&A...382..573F}, making an interaction with the disk surface unlikely. The interaction with the outflow of a second YSO on scales of a few hundred AU only works in binary systems, such as IRS~5, but X-rays are also detected from stars that are apparently single, such as DG~Tau and RY~Tau. A forth possibility is magnetic heating from the reconnection either between two interacting outflows \citep{2008A&A...478..453M} or from fields that are frozen into the outflow \citep{2013A&A...550L...1S}.

In this article, we will analytically explore how a recollimation shock can explain stationary X-ray emission from jets. Based on the observational evidence for multi-layered jets, we assume that the innermost component of the flow is a stellar wind, which is collimated by the disk wind. Collimation shocks of this kind have not been treated in detail in the literature, while X-ray emission due to a moving shock has been studied by several authors \citep[see, e.g.\ the analytical work and numerical simulations by][]{2002ApJ...576L.149R,2007A&A...462..645B}.

\citet{2010A&A...511A..42B} presented simulations with a time variable launching speed, where blobs of material are emitted into the jet every few months or years. Their jet has a radial velocity profile that avoids the growth of random perturbations at the jet boundary and that is compatible with the expected magnetic fields in the environment of young stars. In these simulations faster material catches up with slower, previously emitted matter and shocks form that travel along the jet. Since material is launched almost continuously, the first interaction often takes place fairly close to the star and the simulations show an X-ray emission region only about 100~AU from the star, which fluctuates in luminosity but is present at all times. This region represents only a small fraction of the total simulated X-ray emission from the jet \citep{2010A&A...517A..68B}.

\citet{2011ApJ...737...54B} numerically simulated stationary X-ray shocks. To do so, they impose a rigid nozzle with a radius between 15 and 200 AU and inject a flow of plasma with an intially flat velocity and density profile along the jet axis. The region that accelerates the mass at the bottom of the nozzle is not part of the model, but given the large radius, both disk wind and stellar wind might contribute in such a scenario. \citet{2011ApJ...737...54B} find that a denser layer forms on the walls of the nozzle and that this perturbation travels inward. When this feature reaches the axis of symmetry a diamond-shaped shock forms at a height of 200-300 AU for a nozzle with a radius of 100~AU with temperatures high enough to explain the X-ray emission from HH~154. This model has not been applied to DG Tau, but might provide a viable explanation for the emission in DG Tau, too, if the shape and size of the nozzle is tuned properly.

In contrast to that work, we do not impose rigid boundaries that collimate the flow, but instead prescribe an external pressure profile and then calculate the position of the boundary between the inner wind and the external medium. The setup of \citet{2011ApJ...737...54B} is well-suited to study regions at a distance of 200-300~AU from the star, but in this article we concentrate on the inner region, where the outflow is not yet parallel to the jet axis and stellar and disk outflows have different velocities. Thus, we start with a spherical flow from the stellar surface and explain how a shock can be caused by the recollimation of the stellar outflow due to pressure from the outer disk winds. 
\citet{2012MNRAS.422.2282K} developed a model for this geometry in the context of relativistic jets. Here we apply this model to stellar jets.

\section{The model}
\label{sect:model}
In this section we develop an analytical steady-state model for the interface between the stellar wind and the surrounding disk wind. The pressure of the disk wind collimates the stellar wind into a jet (see figure~\ref{fig:sketch} for a sketch). The two flows are separated by a contact discontinuity, whose exact position is given by pressure equilibrium between the outer, disk wind component and the inner, stellar wind component. As the stellar wind encounters the contact discontinuity, the velocity component perpendicular to the discontinuity is shocked. Thus, our model needs to distinguish three zones: (i) the cold pre-shock stellar wind, (ii) the hot post-shock stellar wind, and (iii) the disk wind. Our goal is to calculate the geometrical shape of the stellar wind shock, since this determines the velocity jump across the shock front and the temperature of the post-shock plasma. 

\subsection{The shape of the shock front}

The Rankine-Hugoniot jump conditions relate the mass density $\rho$, velocity $v$, and pressure $P$ on both sides of a shock. For ideal gases and non-oblique shocks the conservation of mass, momentum and energy across the shock can be written as follows \citep[][chap.~7]{1967pswh.book.....Z}, where the state before the shock front is marked by the index 0, that behind the shock by index 1:
\begin{eqnarray}
\rho_0 v_0 & = & \rho_1 v_1 \label{eqn:RH1}\\
\label{eqn:RH2}P_0+\rho_0 v_0^2 & = & P_1+\rho_1 v_1^2\\
\label{eqn:RH3}\frac{5 P_0}{2\rho_0}+\frac{v_0^2}{2}& = &\frac{5 P_1}{2\rho_1}+\frac{v_1^2}{2} \ .
\end{eqnarray}

We assume that the stellar wind before the shock front is relatively cool (this assumption is justified in Section~\ref{sect:T_0}) and thus the thermodynamic pressure can be neglected, setting $P_0=0$.
The shock front settles at a position where the pressure of the stellar wind equals the post-shock pressure, which in turn determines the position of the contact discontinuity, such that the post-shock pressure equals the confining external pressure of the disk wind $P(z)$ . 

\begin{figure}[h!]
\begin{center}
%\plotone{sketch}
\includegraphics[width=5cm]{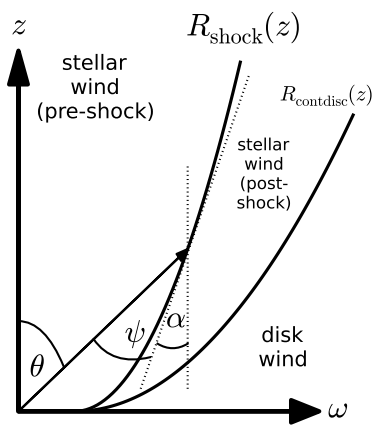}
\caption{\label{fig:sketch}
Geometry in the $(z, \omega)$ plane. The $z$-axis is oriented along the jet, the $\omega$-axis in the plane of the circumstellar accretion disk. The innermost outflow layer is the unshocked stellar wind. The shock front is located at $R_{\rm{shock}}(z)$. The next layer is the hot, post-shock stellar wind, which is separated from the disk wind by the contact discontinuity at $R_{\rm{contdisc}}(z)$. The same model is employed in the paper by \citet{2012MNRAS.422.2282K} and we refer to that publication for a more detailed figure depicting this model.}
\end{center}
\end{figure}

In our case we are dealing with an oblique shock (Figure~\ref{fig:sketch}). Equations~\ref{eqn:RH1} to \ref{eqn:RH3} stay valid if only the velocity component perpendicular to the shock front is taken as $v_0$. 

We use a cylindrical coordinate system $(z, \omega, \theta)$ with an origin at the central star. We place the $z$-axis along the jet outflow direction and assume rotational symmetry around the jet axis. Thus, the flow can effectively be written in $(z,\omega)$. The symbol $r$ denotes the spherical radius, i.e.\ the distance of any point to the star at the origin of the coordinate system. 
In this article we adapt the model of \citet{2012MNRAS.422.2282K} to non-relativistic speeds. Figure~1 in their publication shows the geometry of this model in much detail and we refer to their discussion and their figure~1 for a more detailed description. Although the basic model is the same, we chose to include a similar figure (our Fig.~\ref{fig:sketch}) in this work for the benefit of readers who are not familiar with the work of \citet{2012MNRAS.422.2282K} on extra-galactic jets.

We treat the disk wind as an outer boundary condition with a given pressure profile and concentrate on the description of the stellar wind. To simplify the equations we adopt Kompaneets' approximation \citep{1960SPhD....5...46K} which states that there is no axial pressure gradient so that the pressure profile of the disk wind extends through all layers of the outflow:
\begin{equation}
P(z, \omega, \theta) = P(z)\,.
\end{equation}
With this we can write:
\begin{equation}\label{eqn:Pofz}
\rho_0 v_0^2 = P_{\textrm{post-shock}} = P(z)
\end{equation}

To derive the position of the shock front in the $(z, \omega)$ plane where the pre-shock ram pressure of the stellar wind and the post-shock pressure equal the external pressure $P(z)$, we need to calculate the pre-shock density $\rho_0$ and the pre-shock velocity perpendicular to the shock front $v_0$.

We assume a spherically symmetric stellar wind that is accelerated to its final velocity $v_{\infty}$ within a few stellar radii before any interaction takes place. For a given mass loss rate $\dot M$, the wind density at any distance $r$ from the central star is 
\begin{equation}\label{eqn:rho}
\rho(r) = \frac{\dot M}{4 \pi r^2 v_{\infty}}\ .
\end{equation}
Figure~\ref{fig:sketch} shows that $v_0$ depends on the position of the shock:
\begin{equation}
\label{eqn:v0}v_0 = v_{\infty} \sin \psi
\end{equation}
with 
\begin{equation}\label{eqn:angle}
\psi+\alpha =  \theta \ .
\end{equation}
Again, Figure~\ref{fig:sketch} shows
\begin{equation}\label{eqn:theta}
\tan\theta = \frac{\omega}{z}
\end{equation}
and the angle $\alpha$ is given by the derivative of the position of the shock front:
\begin{equation}\label{eqn:deriv}
\frac{\rm{d}\omega}{\rm{d}z} = \frac{\sin \alpha}{\cos \alpha} = \tan{\alpha}
\end{equation}
This gives:
\begin{equation}\label{eqn:psi}
\psi = \arctan{\frac{\omega}{z}} - \arctan{\frac{\rm{d}\omega}{\rm{d}z}}\ .
\end{equation}
Inserting equation~\ref{eqn:rho} and \ref{eqn:v0} into eqn.~\ref{eqn:Pofz}, we arrive at: 
\begin{equation}\label{eqn:P}
P(z) = \rho_0 v_0^2 = \frac{\dot{M}}{4\pi v_{\infty}(z^2+\omega^2)} v_{\infty}^2 \sin^2(\psi)
\end{equation}
Inserting eqn.~\ref{eqn:psi} this gives an ordinary differential equation (ODE), that describes the shape of the shock front:
\begin{equation}\label{eqn:ode}
\frac{\rm{d}\omega}{\rm{d}z} = \tan\left[\arctan\left(\frac{\omega}{z}\right)-\arcsin\left(\frac{\sqrt{z^2+\omega^2}}{R_0}\right)\right]
\end{equation}
with
\begin{equation}\label{eqn:r0}
R_0(z) = \sqrt{\frac{\dot{M} v_{\infty}}{4\pi P(z)}},
\end{equation}
where $R_0(z)$ is the maximal cylindrical radius of the shock front.

The solution to the ODE determines the location of the shock front. This allows us to calculate the pre-shock velocity perpendicular to the shock front using eqn.~\ref{eqn:v0} and the post-shock temperature $T_{\mathrm{post-shock}}$. From eqn.~\ref{eqn:RH3} with negligible pre-shock pressure (this assumption is justified in Section~\ref{sect:T_0}) and $v_0=4\;v_1$ for a strong shock (the assumption for a strong shock is justified in Section~\ref{sect:T_0}) we derive:
\begin{equation}
T_{\mathrm{post-shock}}(z) = \frac{3}{16} \frac{\mu m_{\textrm{H}}}{k} v_0(z)^2,\label{eqn:T}
\end{equation}
where $m_{\textrm{H}}$ denotes the mass of the hydrogen atom, $k$ the Boltzmann constant and $\mu=0.7$ the mean particle mass for a highly ionized plasma. For general $P(z)$ the ODE needs to be solved numerically\footnote{It is possible to remove all trigonometric functions from eqn.~\ref{eqn:ode} by means of addition formulae, but that introduces singularities into the solution. Thus, we numerically solve the ODE in the form of eqn.~\ref{eqn:ode}.}. This is done in an IPython notebook \citep{PER-GRA:2007}. All code is available at \url{https://github.com/hamogu/RecollimationXrayCTTS/}.

\subsection{Justification of model assumptions}
\label{sect:modelassumptions}
In this section we explain the assumptions made in the derivation of the equations above.

\subsubsection{Sound speed}
\label{sect:soundspeed}
Observations of jets and winds from CTTS indicate that typical temperatures are a few thousand K (except in shocked regions) and typical densities are in the range $10^4-10^6 \mathrm{ cm}^{-3}$ in the optically visible component \citep[e.g.][]{2000A&A...356L..41L,2007ApJ...657..897K}. This might not be the same outflow component that our model describes, but it is the best observational estimate. With those numbers the sound speed $c_s$ is
\begin{equation}
c_s = \sqrt{\gamma \frac{k T}{m_{\textrm{H}}}} \approx 10 \textrm{ km s}^{-1} \; ,
\end{equation}
which is low enough that a strong shock forms even for small $\psi$.

\subsubsection{Initial wind temperature}
\label{sect:T_0}
In Section~\ref{sect:model} we assume an initially cold stellar wind where the thermal pressure before the shock can be ignored. This is motivated by two arguments: (i) \citet{2007IAUS..243..299M} show that {\it hot} stellar winds will cool quickly and cause X-ray emission much brighter than observed if they have a mass loss rate above $10^{-11}M_\odot\mathrm{ yr}^{-1}$. Thus, only stellar winds that are launched {\it cold} can provide the required mass loss rates well above this value (see Section~\ref{sect:masslossrates}). (ii) The solar wind has 1~MK. It is unclear which physical process heats it to this temperature, but it is probably related to magnetic waves. In CTTS the wind mass loss rate is much higher than in the Sun, and it seems likely winds from CTTS are cooler than the solar wind, even if they are heated by the same process.

The approximation to neglect the initial wind temperature is valid to a few $10^5$~K for the densities and wind speeds considered here. Hotter winds cannot be described in our model, because their sound speed is so large that no strong shock develops for small angles $\psi$.

\subsubsection{Magnetic fields}
Our model does not describe magnetic fields.
Two different regions need to be distinguished where magnetic fields can play a role. First, a magnetic field can provide additional pressure in the disk wind and thus contribute to the external pressure $P(z)$. Indeed, the region covered by our model is expected to be inside the Alv\`en surface of the disk wind, so $P(z)$ is probably magnetically dominated (further discussion and references in Section~\ref{sect:boundary}), but for our model only the total value of $P(z)$ matters, independent of the processes that contribute to the pressure. 

Second, the stellar wind could be threaded by a stellar magnetic field. Fields on YSOs are often quite complex with a mixture of open and closed field lines \citep[e.g.][]{2011MNRAS.417..472D,2012MNRAS.425.2948D}.
Qualitatively, closed field lines can either fill with coronal plasma or connect to the accretion disk and carry accretion funnels. Only those parts of the stellar surface connected to open field lines can launch a wind. Thus, the total mass loss rate would be reduced compared to a spherical wind. 
As a simple estimate we calculate the magnetic pressure $P_{\textrm{mag}}=\frac{\boldsymbol{B}^2}{8 \pi}$ for a split monopole field with a field strength of 1~kG at $r=R_\odot$ and compare it to the ram presure (eqn.~\ref{eqn:Pofz}). Using the fiducial parameters from Table~\ref{tab:fiducial} the ram presure dominates over the magnetic pressure already at 0.1~AU and since $P_{\textrm{mag}} \propto \boldsymbol{B}^2 \propto r^{-4}$, while $P_{\textrm{ram}} \propto r^{-2}$ (eqn.~\ref{eqn:Pofz} and \ref{eqn:rho}) we can neglect the magnetic pressure of the stellar wind for our model which predicts typical radial distances on the AU scale.

\section{Constraints on parameters}
\label{sect:parameters}
In this section we discuss observational and theoretical constraints on boundary conditions and input values for the model, most notably $P(z)$, $\dot M$, $v_\infty$, and $\omega(z=0)$. Table~\ref{tab:fiducial} shows the values we adopt as most likely in the following discussion (fiducial model). We vary the parameters individually to show how each of them affects the solution of the ODE. 

\begin{deluxetable}{ccc}
\tablecaption{\label{tab:fiducial}Values for fiducial model and fit to DG Tau}
\tablehead{\colhead{parameter} & \colhead{fiducial} & \colhead{fit to DG~Tau\tablenotemark{a}}}
\startdata
$v_\infty$ & 600 km s$^{-1}$ & $840\pm25$ km s$^{-1}$\\
$\dot M$ & $10^{-8}\;M_\odot\textrm{yr}^{-1}$ & $(5\pm2)\cdot10^{-10}\;M_\odot\textrm{yr}^{-1}$\\
$\omega_0$ & 0.01 AU & = 0.01 AU\tablenotemark{b}\\
$P(z)$ & $P_\infty+P_0\exp\left(-\frac{z}{h}\right)$ & $=P_\infty+P_0\exp\left(-\frac{z}{h}\right)$\\
$P_0$ & $5\cdot 10^{-4}$ Ba &  $(1.4\pm0.2)\cdot 10^{-5}$ Ba\\
$P_\infty$ & $5\cdot 10^{-6}$ Ba & $=0.01\cdot P_0$\\
$h$ & 2 AU & = 5 AU\tablenotemark{b}\\
\enddata
\tablenotetext{a}{The errors represent statistical uncertainties from the $\chi^2$ fit. Systematic uncertainties are discussed in the text.}
\tablenotetext{b}{fixed during fit}
\end{deluxetable}

\subsection{Disk winds as boundary conditions for stellar winds}
\label{sect:boundary}
Different models exist to explain wind launching from the stellar surface \citep{1988ApJ...332L..41K,2005ApJ...632L.135M}, the X-point close to the inner disk edge \citep{1994ApJ...429..781S} and magneto-centrifugal launching from the disk \citep{1982MNRAS.199..883B,2005ApJ...630..945A}. It is likely that more than one mechanism contributes to the total outflow from the system. In this case, we expect a contact discontinuity between the different components whose position is determined by the pressure on both sides. Specifically, hydromagnetic disk winds have a tendency to collimate and possibly even to recollimate to smaller flow radii under certain conditions \citep{1982MNRAS.199..883B,1992ApJ...394..117P}.
Numerically, the magneto-centrifugally accelerated disk wind is probably the best explored component. Magneto-hydrodynamic (MHD) simulations of the disk wind have been performed in 2D \citep[e.g.][]{2005ApJ...630..945A}, 2.5D \citep[e.g.][]{2011ApJ...728L..11R} or 3D \citep[e.g.][]{2006ApJ...653L..33A}, but typically do not resolve the stellar wind. However, they show that the disk wind is collimated close to the axis and that the densities are largest in this region. Furthermore, the Alfv\'en surface (which separates the magnetically dominated region from the flow-dominated region) is located at many AUs for the inner layers of the jet. This is in contrast with the outer, less collimated layers of the wind, which leave the magnetically dominated region at a few AUs.

\citet{2009A&A...502..217M} present analytical and numerical solutions for several scenarios that mix an inner stellar wind and an outer disk wind \citep[this model has been extended in][]{2012A&A...545A..53M,2014A&A...562A.117T}. In contrast to our approach, they impose a smooth transition between stellar wind and disk wind and they start their simulation at $z=50$~AU instead of at the star. With some time variability in the wind launching their models produce knot features in the jet. In the context of our analysis, we note that the pressure in their models is magnetically dominated and that Kompaneet's approximation does not hold in the disk wind, but that the presure gradient is small on scales of a few AU. The pressure at the jet axis is high in the plane of the disk and drops by one to two orders of magnitude until it reaches a plateau ($P_\infty$). Below we use an exponential $P(z)=P_\infty+P_0\exp\left(-\frac{z}{h}\right)$ to mimic this profile.
Similar profiles for the inner density and pressure are seen in simulations by other groups \citep[e.g.][]{2005ApJ...630..945A,2006ApJ...653L..33A,2008ApJ...678.1109M}.

\begin{figure*}[h!]
\begin{center}
\plotone{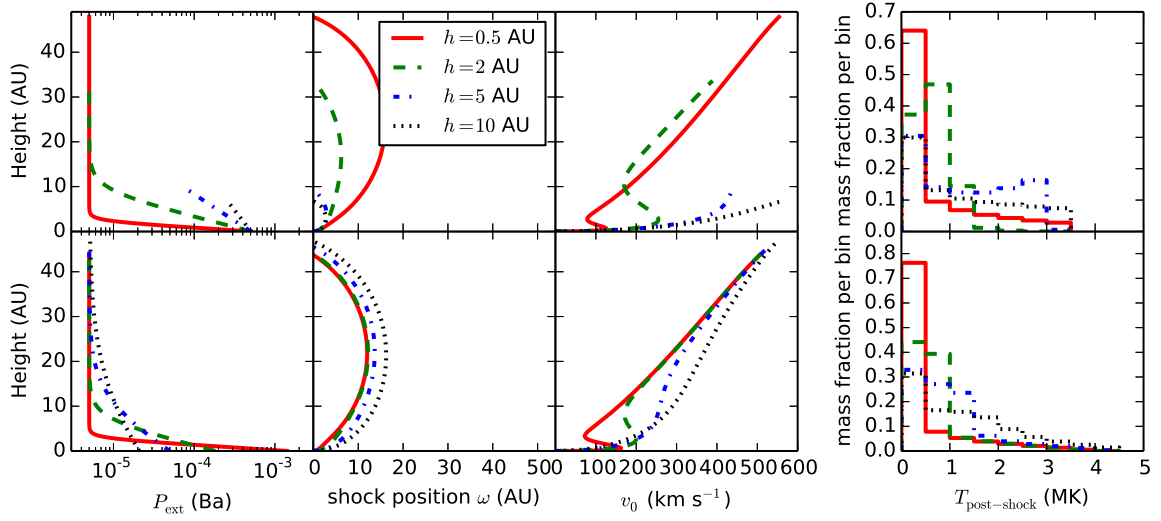}
\caption{\label{fig:p_ext}
Solutions to the ODE for different pressure profiles. In the top row the scale height $h$ varies while all other parameters are fixed; the bottom row uses the same values for $h$, but also scales $P_0\propto h^{-1.5}$.
\emph{leftmost panel}: Profile of the external pressure $P(z)$. 
\emph{middle left panel}: Position of the shock front. Note that the two axes in this panel use the same scale. The unshocked stellar wind region is much longer in $z$ direction (height above the disk) than it is wide. 
\emph{middle right panel}: Pre-shock velocity $v_0$ measured perperdicular to the shock front. 
\emph{rightmost panel}: Distribution of post-shock temperatures, weighted by spherically integrated mass flux. This distribution is dominated by the temperature (which in turn is set by $v_0$) at small values of $z$, since the shock surface is close to the central object and covers a large solid angle of the stellar wind and therefore a large fraction of the total mass flux. Note that the histograms show the temperature directly behind the shock front and does not account for the fact that all plasma will eventually cool and contribute emission at cooler temperatures; a simulation of the thermodynamics of the cooling plasma is beyond the scope of this article.}
\end{center}
\end{figure*}

Figure~\ref{fig:p_ext} shows how different pressure profiles influence the shock position. Larger pressures force the shock front onto the symmetry axes for smaller $z$ (top row). 
The solutions shown in the bottom row of the figure are for the same pressure scale heights $h$ as those in the upper row, but here we use smaller $P_0$ for scenarios with large $h$, so that the shock front reaches the jet axis at approximately the same $z$. Close to the disk plane the pre-shock speeds differ significantly, but at large $z$ they reach very similar values. The scenarios with smaller $P_0$ reach larger radii and the slightly different shape of the shock front leads to more plasma at high temperatures.

\subsection{Mass-loss rates}
\label{sect:masslossrates}
The measured mass loss rates in the outflows from CTTS vary widely between objects.  Even for a single object, very different mass loss rates can be found, depending on the spectral tracers chosen and on the assumptions used to calculate mass loss rates from line fluxes. The filling factor that describes the fraction of the observed volume occupied by hot gas is especially uncertain because the innermost jet component is generally not resolved.

Typical mass loss rates found in the literature for CTTS outflows are in the range $10^{-10}-10^{-6}M_{\odot}\textrm{ yr}^{-1}$ \citep{1999A&A...342..717B,2006A&A...456..189P}. \citet{2006ApJ...646..319E} measure values down to $10^{-10}$~M$_{\odot}$~yr$^{-1}$ for some CTTS, but only upper limits for weak-line T Tauri stars (WTTS). In the specific case of DG~Tau \citet{1997A&A...327..671L} calculate the  mass loss rate as $6.5\cdot 10^{-6}$~M$_{\odot}$~yr$^{-1}$; \citet{1995ApJ...452..736H}
obtain $3\cdot 10^{-7}$~M$_{\odot}$~yr$^{-1}$ and, further out in the jet, \citet{2000A&A...356L..41L} find $1.4\cdot 10^{-8}$~M$_{\odot}$~yr$^{-1}$. Those measurements for the optical jet are probably dominated by the disk wind \citep[e.g.][]{2014arXiv1404.0728W} and unlikely to track the stellar mass loss correctly.
\citet{2009A&A...493..579G} show that a mass loss below $10^{-10}$~M$_{\odot}$~yr$^{-1}$ is sufficient to explain the X-ray emission from the jet as shock heating.
We use $10^{-8}$~M$_{\odot}$~yr$^{-1}$ as fiducial stellar mass loss in the remainder of the article. This is only a fraction to the total mass loss of the system because the disk wind, though slower, operates over a much larger area and dominates the system's mass loss.

\begin{figure*}[h!]
\begin{center}
\plotone{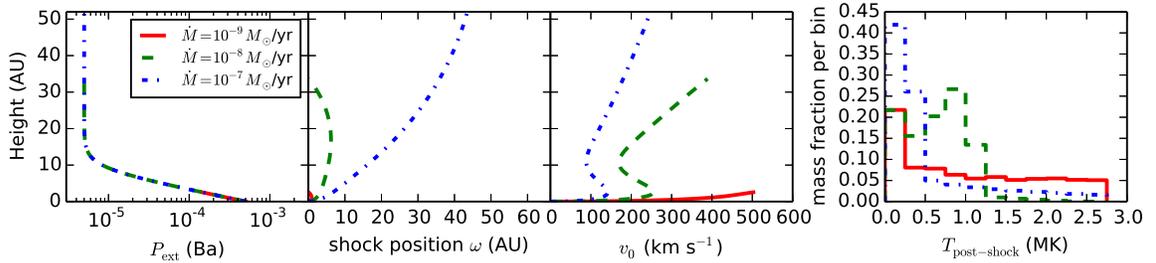}
\caption{\label{fig:dot_m}
Solutions to the ODE for different mass loss rates. For a description of the individual panels see Fig.~\ref{fig:p_ext}. The middle left panel only shows the region close to the star for clarity. The shock front with the largest mass loss rate extends to $\omega=50$~AU at $z=60$~AU and comes back to the axis of symmetry at 130~AU.}
\end{center}
\end{figure*}

Figure~\ref{fig:dot_m} shows how a larger mass loss rate and therefore a higher density and ram pressure in the stellar wind pushes the shock front out to larger radii. The different shape of the shock front also influences the post-shock temperatures. In the high mass loss rate scenario (blue dash-dotted line) the shock front reaches its maximum radius at a height of 60~AU and most of the spherically symmetric wind passes the shock front at shallow angles, so this scenario has the highest fraction of low temperature material.

\subsection{Wind speed}
The launching mechanism of the stellar wind in CTTS is uncertain. \citet{2007IAUS..243..299M} show that stellar winds from CTTS cannot have a total mass loss above $10^{-11}M_\odot\mathrm{ yr}^{-1}$ if they are launched hot. 
Thus, the winds of CTTS are probably more complex than just a scaled up version of the solar wind.
Still, launching velocities similar to the solar wind seem to be a reasonable estimate for $v_\infty$. A velocity of a few hundred km~s$^{-1}$ is compatible with the fastest speeds observed from optical line shifts in jets (see references in Section~\ref{sect:introjetobs}). It also matches the depth of the gravitational potential, and while the launching mechanism is unknown, it is likely powered by energy released in the accretion process \citep{1988ApJ...332L..41K,2005ApJ...632L.135M}.

The solar wind consists of a slow wind with a typical velocity of 400~km~s$^{-1}$ and a fast wind around 750~km~s$^{-1}$ \citep{2005JGRA..110.7109F}. The relative contribution and the launching position of the two types changes over the solar cycle, but the slow wind often emerges from regions near the solar equator and the fast wind is generally associated with coronal holes \citep{1999GeoRL..26.2901G,2003A&A...408.1165B,2009LRSP....6....3C}. Despite these differences, the total energy flux in the solar wind is almost independent of the latitude, because the slower wind is denser than the faster wind \citep{2012SoPh..279..197L}. We set $v_\infty=600$~km~s$^{-1}$ as the fiducial outflow velocity and we assume that the wind is accelerated close to the star and has reached $v_\infty$ before it interacts with the shock front. We use a spherically symmetric stellar wind with a constant velocity. For a solar-type wind this works well for deriving the shape of the shock front because eqn.~\ref{eqn:r0} depends only on the total energy flux $\rho v^2_\infty \propto \dot M v_\infty$ and not the velocity itself. 

\begin{figure*}[h!]
\begin{center}
\plotone{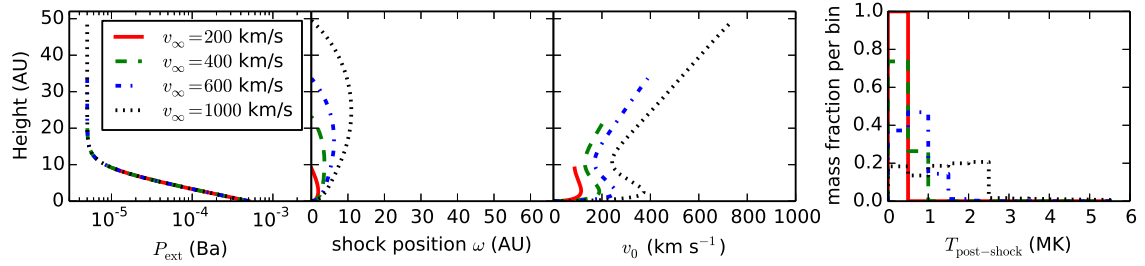}
\caption{\label{fig:v_infty}
Solutions to the ODE for different $v_\infty$. For a description of the individual panels see Fig.~\ref{fig:p_ext}.}
\end{center}
\end{figure*}

Figure~\ref{fig:v_infty} shows how a large $v_\infty$ and a correspondingly large ram pressure push the shock front higher above the disk plane, similar to outflows with a larger $\dot M$. Additionally, $v_\infty$ is the most important parameter that controls the post-shock temperatures.
The figure shows that high shock speeds and thus high post-shock temperatures are reached close to the disk plane. Because this region covers a large solid angle, it is an important contributor to the total temperature distribution of the post-shock plasma (rightmost panel). However, the temperature in this region is probably overestimated by our model because the wind may not have reached $v_\infty$ this close to the star or the wind may not be spherically symmetric. A lower wind velocity and higher density at the equator similar to the solar low-velocity wind would lead to lower post-shock temperatures with a higher emission measure close to the disk plane.

\subsection{Starting point of integration}
\label{sect:omega0}
\begin{figure*}[h!]
\begin{center}
\plotone{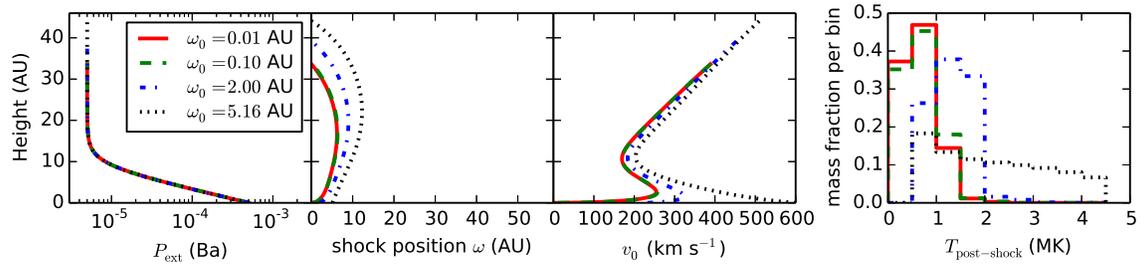}
\caption{\label{fig:omega_0}
Four solutions to the ODE for four different starting points $\omega(z=0)=\omega_0$. For a description of the individual panels see Fig.~\ref{fig:p_ext}.}
\end{center}
\end{figure*}

From a mathematical point of view, the starting point of the integration in the plane of the disk can be chosen freely anywhere between $\omega=0$ and $\omega=R_0(z=0)$. Figure~\ref{fig:omega_0} compares different starting points under otherwise equal conditions. For small initial radii the ram pressure of the stellar wind pushes the shock surface out in a small $\Delta z$. This leads to small pre-shock speeds in this region because the direction of the flow and the shock surface are almost parallel. This region also represents a large fraction of the total mass loss of the stellar wind, because it covers a large solid angle. Consequently, models with small values of $\omega_0$ heat less material to high temperatures. 

Physically, the position of the shock front is restricted by the position of the disk - the shock between the stellar wind and the disk material (in the disk itself or the disk wind) must occur within the inner hole of the disk. Figure~\ref{fig:omega_0} shows that the two solutions for $\omega_0=0.01$~AU and $0.1$~AU are almost indistinguishable and the exact value for this parameter is not important as long as $\omega < 0.5$~AU. We use $\omega_0 = 0.01$~AU as the fiducial starting point for the integration.

\section{Results}
\label{sect:results}

\subsection{Uncertainties}
All input parameters discussed above can take a range of values. We argue that the starting point of the integration has to be within the inner disk and the exact value has only a small influence on the result. The wind velocity is closely related to the temperature of the observed plasma, the mass loss rate to the size of the region and the luminosity.

The biggest uncertainty is the value of the external pressure. As discussed above, different simulations in the literature predict similar pressure profiles, but the normalization of the pressure depends to a large degree on the disk magnetic field, which is only poorly constrained. In our calculation, we have scaled the pressure such that the post-shock densities are compatible with observations of the jet.

\subsection{Size of stellar wind zone}
For all parameters consistent with the theoretical and observational constraints the stellar wind is enclosed in a finite region by a shock front. This shock front generally reaches a maximum cylindrical radius of a few AUs, and a much larger height above the accretion disk for external pressure profiles with high pressure in the plane of the disk and a large pressure gradient (fiducial model in Fig.~\ref{fig:result}). A shallower pressure profile leads to a stellar wind region that is wider. 

\subsection{X-ray luminosities}
\label{sect:LX}
The post-shock plasma is less dense than the typical stellar corona and can thus be treated in the so-called coronal approximation, meaning that the plasma is optically thin and line ratios for prominent X-ray lines are in the low-density limit. We use the shock models of \citet{2007A&A...466.1111G} to predict the fraction of the total pre-shock kinetic energy that will be emitted in the X-ray range. We refer to that publication for details on the shock models. In summary, the models simulate radiative cooling of optically thin plasma in a two-fluid approximation, where electrons and ions are described with a Maxwellian velocity distribution, each with a different temperature. The ionization and recombination rates are calculated explicitly, but it turns out that the inonization state differs from the ionization equilibrium only in a small fraction of the post-shock cooling zone, even for densities as low as $10^5$~cm$^{-3}$.

\citet{2011AN....332..448G} published a grid of X-ray spectra\footnote{Available at http://hdl.handle.net/10904/10202} based on these models with pre-shock velocities between 300 and 1000~km~s$^{-1}$ in increments of 100~km~s$^{-1}$. We integrate all emission between 0.3 and 3~keV for each spectrum. At 300~km~s$^{-1}$ only 2\% of the available energy is emitted between 0.3 and 3~keV (Figure~\ref{fig:fracxray}), so we set the fraction to zero for pre-shock velocities of 0, 100 and 200~km~s$^{-1}$, which are not covered by the model grid. The fraction of energy emitted in X-rays is independent of the density except for a few density-sensitive emission lines with negligible contribution to the integrated flux. The physical size of the post-shock region depends strongly on the density, but total energy available only depends on the pre-shock velocity and the total mass flux. Thus, the X-ray luminosity $L_X$ does not change, if the post-shock region is compressed by some external pressure.

The highest post-shock temperatures are generally reached at the base of the jet when the stellar wind encounters the inner disk rim or at large $z$ when the shock front intersects the jet axis. In our fiducial model (Fig.~\ref{fig:result}, solid red line), the pre-shock velocity is $>250$~km~s$^{-1}$ at $z<5$~AU and $z>20$~AU. Given the large solid angle covered by the inner disk rim, the  $z<5$~AU region contributes significantly to the total mass flux (compare the red line and the red filled histogram in Figure~\ref{fig:result}, rightmost panel). However, in most YSOs the central object is highly absorbed. Therefore, we calculate all $L_X$ values taking into account only regions with  $z>5$~AU. For the fiducial, the high $v_\infty$, the low $\dot M$, and the shallow $P$ model in Figure~\ref{fig:result} the predicted $L_X$ is $3\cdot10^{29}$, $5\cdot10^{30}$, $1\cdot10^{28}$, and $1\cdot10^{31}$~erg~s$^{-1}$, respectively.
\citet{2009A&A...493..579G} already showed that in DG~Tau a small fraction, about $10^{-3}$, of the total mass loss rate in the outflow is enough to power the observed X-ray emission at the base of the jet. In our fiducial model, this small fraction corresponds to the mass flow close to the jet axis, where the pre-shock velocities are highest.

\subsection{The size of the post-shock zone}
\begin{figure*}[h!]
\begin{center}
\plotone{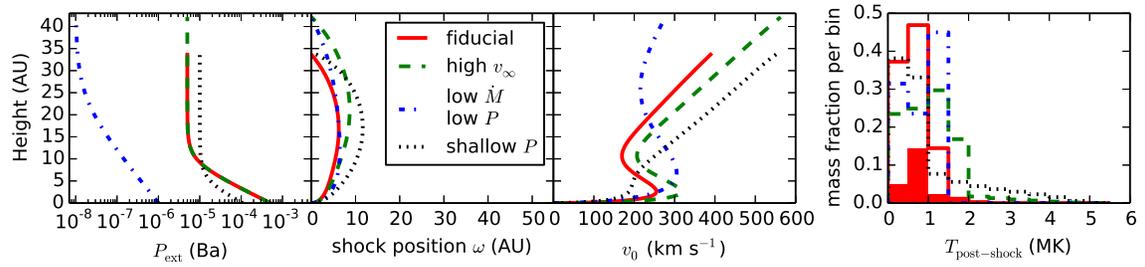}
\caption{\label{fig:result}
Different scenerios with parameters well within the range observed for CTTS predict a stellar wind shock that reaches 20-40~AU along the jet, but extends only few AU in the radial direction. For a description of the individual panels see Fig.~\ref{fig:p_ext}. The solid red histogram in the rightmost panel shows the post-shock temperature distribution for the fiducial model for $z>5$~AU, while the red solid lines shows the temperature distribution for all $z$ for the same model. All plasma with $T_{\mathrm{post-shock}}>1.5$~MK is found at $z>5$, but only a small fraction of the cooler plasma.}
\end{center}
\end{figure*}

\begin{figure}[h!]
\begin{center}
\plotone{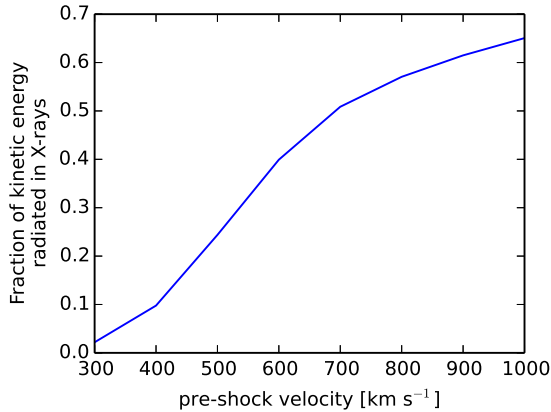}
\caption{\label{fig:fracxray}
Fraction of the kinetic energy of a shock that is radiated between 0.3 and 3.0~keV according to the shock models of \protect{\citet{2007A&A...466.1111G}}. A small dependence on the pre-shock density is ignored.
}
\end{center}
\end{figure}

\begin{figure}[h!]
\begin{center}
\plotone{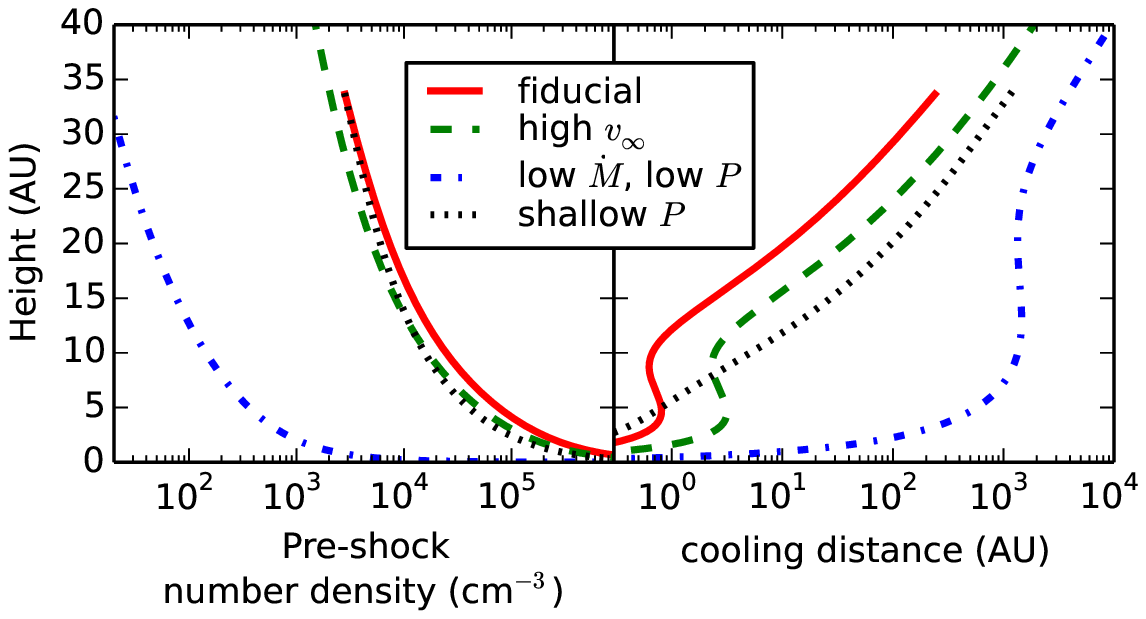}
\caption{\label{fig:rhocool}
Pre-shock number density and estimate of the cooling distance (see text) for the shock models shown in Fig.~\ref{fig:result}.}
\end{center}
\end{figure}

Figure~\ref{fig:rhocool} shows the pre-shock number densities $n_0$ for the four models from Fig.~\ref{fig:result}. A detailed treatment of the post-shock region is beyond the scope of this paper, but an estimate for the post-shock cooling length $d_{\mathrm{cool}}$ can be derived according to \citet{2002ApJ...576L.149R}:
\begin{equation} \label{eqn:dcool}
d_{\mathrm{cool}} \approx 20.9 \mathrm{ AU}
    \left(\frac{10^5\mathrm{ cm}^{-3}}{n_0}\right)
    \left(\frac{v_{\mathrm{shock}}}{500\mathrm{ km s}^{-1}}\right)^{4.5}\ .
\end{equation}
The derivation for this formula assumes a cylindrical cooling flow. In contrast, in our model the external pressure can change with $z$ and might compress the post-shock gas or allow it to expand radially. Since denser gas emits more radiation and thus cools faster, $d_{\mathrm{cool}}$ is only an estimate. However, for our models $P(z)$ has already reached $P_\infty$ at the position of the shock and thus eqn.~\ref{eqn:dcool} is valid. With this in mind, figure~\ref{fig:rhocool} (right panel) indicates that the cooling lengths for our fiducial model is consistent with the X-ray observations that do not resolve the wind shock \citep{2008A&A...488L..13S}.

\subsection{Fit to DG Tau data}
In this section we perform a formal fit of our model to the \emph{Chandra} data of DG~Tau. 
Given the limitations of the model, there are some systematic uncertainties in the fitted values. Nevertheless the model reproduces the observations with parameters that are in line with the observational limits discussed above and thus shows that our model is consistent with the data.

We download and extract the following \emph{Chandra} observations from the archive: ObsIDs \dataset[ADS/Sa.CXO#obs/04487]{4487}, \dataset[ADS/Sa.CXO#obs/06409]{6409}, \dataset[ADS/Sa.CXO#obs/07246]{7246}, and \dataset[ADS/Sa.CXO#obs/07247]{7247}. These datasets have been used and are described in \citet{2008A&A...478..797G}, \citet{2008A&A...488L..13S}, and \citet{2009A&A...493..579G} and we refer the reader to those publications for more details. We processed the data with CIAO~4.6 \citep{2006SPIE.6270E..60F}, extracting the CCD spectrum from DG~Tau and a larger, source-free background region on the same chip with the \texttt{specextract} script.  \citet{2008A&A...478..797G} showed that the spectral properties of the soft X-ray component are compatible within the errors in all four observations and \citet{2008A&A...488L..13S} demonstrate that the offsets measured between the soft and the hard component is compatible as well. 
Because the number of counts in the soft component is low and we are not interested in the rapid changes seen in the hard emission attributed to the stellar corona \citep{2008A&A...478..797G} we combine all four source spectra. We bin them to 25 counts per bin and subtract the background. We fit a model with two thermal optically thin plasma emission components \citep[APEC,][]{2012ApJ...756..128F} each with its own cold absorber analogous to \citet{2008A&A...478..797G}. We then replace the cooler APEC component by our recollimation shock model. Since the numerical evaluation of this model is slow, we fix the properties of the hot, stellar component at the values obtained in the previous fit to reduce the number of free parameters. To further reduce the number of parameters, we set $P_0 = 100\times P_\infty$ and $h=5$~AU. For each set of parameters, our model is evaluated as follows: We solve the ODE in eqn.~\ref{eqn:ode} numerically and calculate mass flux and pre-shock velocity for each numerical step. To take the high absorbing column density to DG~Tau itself into account we discard all steps with $z<5$~AU. We bin the remaining mass flux according to the pre-shock velocity in bins of 250-350, 350-450, ..., 950-1050~km~s$^{-1}$. For each bin we select the appropriate post-shock cooling spectrum from the model grid discussed in Section~\ref{sect:LX} and scale it with the mass flux and an assumed distance to DG~Tau of 140~pc \citep{1994AJ....108.1872K}. We use the Sherpa fitting tool \citep{2001SPIE.4477...76F} to adjust $\dot M$, $P_0$, $v_\infty$, and the absorbing column density $N_\textrm{H}$ along the line-of-sight to the shock to simultaneously reproduce the observed X-ray spectrum and the position of the shock. \citet{2008A&A...488L..13S} and \citet{2011ASPC..448..617G} give distances of 25-45~AU between DG~Tau and the soft X-ray emission, but do not calculate formal errors for the position. For the purpose of a $\chi^2$ fit we compare the position where the shock front intersects the jet axis to the value $z_{max} = 30\pm5$~AU.

\begin{figure}[h!]
\begin{center}
\plotone{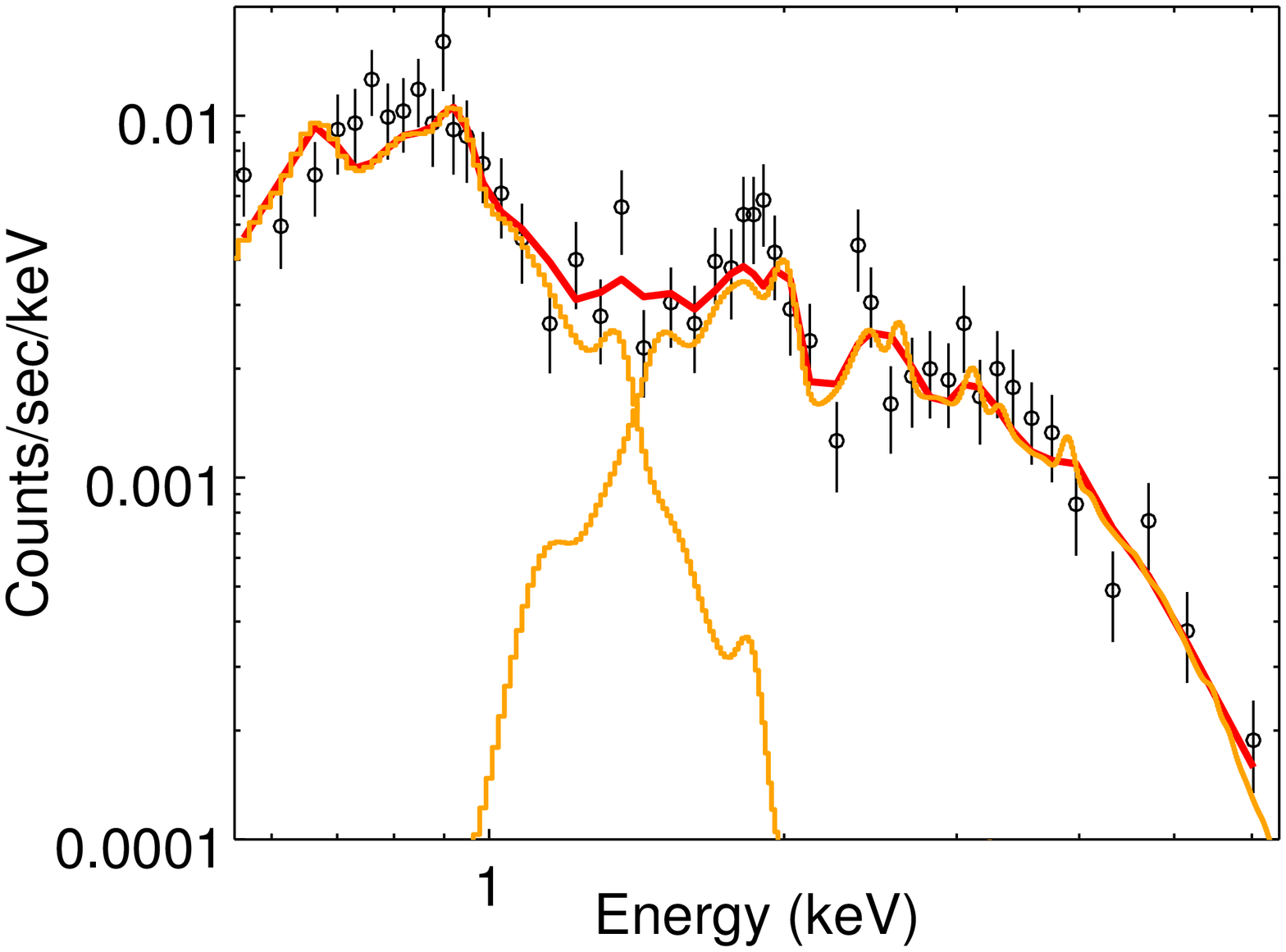}
\caption{\label{fig:fit}
Observed X-ray spectrum of DG~Tau. The best-fit model is shown in red; orange lines represent the weakly absorbed recollimation shock (dominating at low energies) and the strongly absorbed thermal emission component (dominating at high energies).}
\end{center}
\end{figure}

The best-fit parameters for the shock model are given in table~\ref{tab:fiducial}, $N_\textrm{H}=(4.7\pm0.3)\cdot10^{21}\mathrm{ cm}^{-2}$ for the shock, and the parameters of the hot coronal component are $N_\textrm{H}=2.6\times10^{22}$~cm$^{-2}$, plasma temperature $kT = 2.2$~keV, and volume emission measure $VEM=5\times10^{52}$~cm$^{-5}$. The best-fit has $z_{max} = 30.4$~AU and $\chi^2_{red}= 1.1$. The fitted spectrum is shown in Figure~\ref{fig:fit} where the black dots with error bars represent the data, the red line shows the full model, and the orange lines show the individual model components. The shock and coronal component dominates at low and high energies respectively.

\section{Discussion}
\label{sect:discussion}
We show that stellar wind and disk wind can interact in a CTTS system. The magnetic and thermodynamic pressure of the disk wind can confine the stellar wind into a narrow, jet-like region, bound by an elongated shock surface. For reasonable parameters of $\dot M$, $v_\infty$, $\omega_0$ and $P(z)$ the shock surface encloses a region a few AU wide and tens of AU along the jet axis, but our model makes no statement about emission that originates further out in the jet such as Herbig-Haro knots. Only a fraction of the kinetic energy of the stellar wind is converted into heat in the recollimation shock and the remaining velocity can still be sufficient to heat the jet material again when it encounters another obstacle, such as the ISM.

Most of the imaging of YSO winds traces molecular lines and low-ionization stages, e.g. \ion{O}{1} or \ion{Fe}{2}. These lines are formed in low-temperature regions, but not in a hot post-shock plasma. Thus, one could expect to see a hole that is filled by hot post-shock plasma from the stellar wind. However, no such hole is resolved in any CTTS imaging. Our calculations show that the shock surface is so small that it cannot be resolved with current instrumentation\footnote{HST imaging and AO corrected, ground-based IR observations reach a resolution around 0\farcs1, which corresponds to 15~AU for DG Tau -- one of the closest YSO jets. However, saturation or coverage by a coronagraphic disk often mean that even structures slightly larger can be missed in images, if they are located very close to the central star.} and therefore cannot be seen directly as a cavity in the disk wind. A small fraction of the stellar wind is shocked to X-ray emitting temperatures $>1$~MK and provides a stationary X-ray source consistent with observations. 
We show that such a shock naturally arises in a scenario where the stellar wind feeds the innermost layer of the jet because it is confined by external pressure.
Furthermore, \citet{2013A&A...550L...1S} observed C\;{\sc iv} emission in DG Tau that is formed at cooler temperatures than those required for X-ray emission. In recent observations in the IR \citet{2014arXiv1404.0728W} also identified a stationary emission region on the jet axis about 40~AU from the central star. They interpret the X-rays,  C\;{\sc iv}, and their own [ Fe\;{\sc ii}] data all as a signature of the same shocked jet, while \citet{2013A&A...550L...1S} point out that the C\;{\sc iv} luminosity is too large to be powered by just the cooling X-ray plasma. Looking at the post-shock temperature distribution in Fig.~\ref{fig:result}, our model can naturally explain multiple temperature components in the stellar wind.

However, more detailed numerical simulations of the post-shock cooling zone and the shape of the contact discontinuity between the disk wind and the post-shock stellar wind are required to check whether the physical extent of the cooling region behind the shock front and the position of the peak  C\;{\sc iv} emission can be matched to the observations.

In any case, Figure~\ref{fig:fit} shows that the model can explain the observed X-ray spectra. The best-fit values obtained for DG~Tau have a significantly higher velocity and lower mass loss rate than the fiducial model. The parameters for the fiducial model are chosen to match the flow velocities and mass loss rates that are observed in jets. Yet, the higher velocity and lower mass loss rate do not directly contradict those observations, because the pre-shock stellar wind extends only over a small area, so that it presumably contributes little to the luminosity in the optical emission lines compared to the inner disk wind. Consequently, the high velocity in the stellar wind is not directly observable. We also note that in the fit we varied only the normalization of the external pressure, but not the spatial profile. Different profiles lead to different shock-front shapes and shock velocities and thus require an adjustment of $v_\infty$ and $\dot M$ to fit the X-ray data. However, the pressure profile caused by the disk wind and disk magnetic field is not very well constrained. Thus, the accuracy of the fitted numbers is not so much limited by the statistical uncertainty given in table~\ref{tab:fiducial}, but by the systematics of the model. The best-fit values should not be taken at face value, but they demonstrate that a recollimation shock is one possible explanation for the observed X-ray emission. 

The idea of a recollimation shock is not new.
\citet{1993ApJ...409..748G} discussed a similar idea as we do here, where they aim to explain the forbidden optical emission lines seen from CTTS with a shock due to the recollimation of the jet outflow. In contrast to our model, they attribute it to the shocked disk wind, not the stellar wind. However, a shocked disk wind cannot supply the high shock velocities to explain the resolved X-ray and \ion{C}{4} emission that we now see. Our model, a shocked stellar wind, is collimated because it is embedded into a strong disk wind. We expect that the low-temperature emission from the stellar wind is small compared to the low-temperature emission from the surrounding disk wind. Only for high temperatures (X-ray and FUV emission), the stellar wind will dominate because it is much faster.

We now compare our work to the simulations of \citet{2010A&A...511A..42B,2010A&A...517A..68B} which use smooth velocity and density profiles for the jet with radii between 8 and 200~AU -- larger than almost all simulations shown in this article.  \citet{2010A&A...517A..68B} find X-ray emission of varying luminosity around 100~AU from the central source in their simulations for the HH~154 jet and potentially different jet parameters can cause this feature to appear closer to the star. However, this emission is much weaker than the X-ray emission at larger scales in contrast to the situation in DG~Tau. Further simulations are required to test if realistic launching conditions can also make a quasi-stationary X-ray shock that outshines the knots at larger distances. Another possibility that can be tested in future simulations is that our picture of a wind-wind interaction can be combined with a time variable launching speed. At distances of only a few AU the stellar wind and the disk wind would interact and cause a stationary collimation shock as explained in this article. After passing through the shock, the stellar wind and the outer disk wind might mix, so that the jet could appear more homogeneous at larger distances. If the intital launching velocity is time variable, not only will the properties of the recollimation shock change, but the velocity of the combined outflow would also vary and could thus cause moving shock fronts further out in the jet as in the simulations of \citet{2010A&A...511A..42B,2010A&A...517A..68B}.
\citet{2011ApJ...737...54B} find a diamond-shaped shock and again it is possible that slightly different jet parameters can cause this feature to appear closer to the star as seen in DG~Tau.

\citet{2009A&A...502..217M,2012A&A...545A..53M} also perform numerical simulations of a jet confined by a disk wind. Their simulations again deal with larger distances from the central star and they concentrate on knots in the jet. Yet, their bubbles of shock heated gas have very similar shapes compared with our results in Figure~\ref{fig:result}. This indicates that this form is robust. 

Does the scenario of a stellar wind recollimation shock as X-ray source also apply to other CTTS or is it specific to DG Tau? While there is no reason to believe that the wind launching in DG~Tau is unique, it certainly presents us with a special viewing geometry, where the star itself is heavily absorbed, but the jet shock at 30~AU is visible. At the same time, the first knot in the jet that shows X-ray emission is located at 700~AU and thus can be clearly separated from the inner, presumably stationary emission. \citet{2011A&A...530A.123S} analyze three epochs of X-ray emission from HH~154 and see an inner stationary component and a variable component at slightly larger radii \citep[see also][]{2011ApJ...737...54B}. However, there is no gap between both components and the small number of photons makes it difficult to quantify variability and proper motion. Also, the emission from HH~154 is more energetic and would require much larger wind velocities if it is due to a stellar wind recollimation shock. In other young stars with resolved X-ray emission, the central star is either visible and outshines any potential recollimation shock (e.g.\ HD~163296 \citep{2005ApJ...628..811S,2013A&A...552A.142G} or RY~Tau \citep{2014ApJ...788..101S}) or embedded so deep into the cloud that a wind shock at a few tens of AU would be completely absorbed \citep[e.g.\ HH80/81][]{2004ApJ...605..259P}. Thus, we cannot decide this question observationally, but it seems reasonable to assume that the same processes that lead to a recollimation shock in the stellar wind in DG~Tau, should also operate in other CTTS, even if less favourable conditions make it harder to observe in X-rays.

\section{Summary and conclusion}
\label{sect:summary}
A fast stellar wind that is confined by an external pressure from the disk wind will form a stationary collimation shock \citep{2012MNRAS.422.2282K}. We derive the geometrical shape and other properties of this shock front using parameters appropriate for young stars and find that it encloses a roughly egg-shaped volume for all combinations of parameters discussed in this article. This model provides a viable explanation of the soft X-ray and FUV emission observed at the base of young stellar jets and thus presents an alternative to other models on the same topic in the literature (see discussion in Section~\ref{sect:intromodel}). Specifically for DG~Tau, we fit the parameters of the stellar wind to the observed X-ray spectrum and the observed distance between star and X-ray emission.  We expect recollimation shocks to be present in other jets, but only in the case of DG~Tau it is possible to test this observationally.

This analysis shows that collimation shocks in a stellar wind are one possible model to explain stationary X-ray emission in the jets of CTTS, but we cannot claim that recollimation shocks of stellar winds are the only possible explanation until all competing scenarios are analyzed to the same level of detail to be ruled out or confirmed. We stress that our scenario makes no statement about the X-ray emission observed beyond the initial recollimation shock.

\acknowledgments
Support for this work was provided for HMG by NASA through grant GO-12907.01-A from the Space Telescope Science Institute, which is operated by the Association of Universities for Research in Astronomy, Inc., under NASA contract NASA 5-26555 and by NSF AST1313083 and NASA NNX14AB38G for ZYL. PCD acknowledges support by the DLR under 50 OR 1307. We thank T.~Matsakos for discussions about his jet simulations and an anonymous referee for suggestions that helped to improve the relevance and clarity of the paper.

\bibliography{biblio}
\end{document}